\documentclass[floatfix,aps,prl,twocolumn,superscriptaddress]{revtex4}
\usepackage{graphicx}
\usepackage[]{epsfig}
\usepackage{times,amsmath,amssymb}

\usepackage{color}

\begin{document}

\title{Difference in charge and spin dynamics in a quantum dot-lead coupled system}

\author{Tomohiro Otsuka}
\email[]{tomohiro.otsuka@riec.tohoku.ac.jp}
\affiliation{Research Institute of Electrical Communication, Tohoku University, 2-1-1 Katahira, Aoba-ku, Sendai 980-8577, Japan}
\affiliation{Center for Emergent Matter Science, RIKEN, 2-1 Hirosawa, Wako, Saitama 351-0198, Japan}
\affiliation{Department of Applied Physics, University of Tokyo, Bunkyo, Tokyo 113-8656, Japan}
\affiliation{JST, PRESTO, 4-1-8 Honcho, Kawaguchi, Saitama 332-0012, Japan}

\author{Takashi Nakajima}
\affiliation{Center for Emergent Matter Science, RIKEN, 2-1 Hirosawa, Wako, Saitama 351-0198, Japan}
\affiliation{Department of Applied Physics, University of Tokyo, Bunkyo, Tokyo 113-8656, Japan}

\author{Matthieu R. Delbecq}
\affiliation{Center for Emergent Matter Science, RIKEN, 2-1 Hirosawa, Wako, Saitama 351-0198, Japan}

\author{Peter Stano}
\affiliation{Center for Emergent Matter Science, RIKEN, 2-1 Hirosawa, Wako, Saitama 351-0198, Japan}
\affiliation{Department of Applied Physics, University of Tokyo, Bunkyo, Tokyo 113-8656, Japan}
\affiliation{Institute of Physics, Slovak Academy of Sciences, 845 11 Bratislava, Slovakia}

\author{Shinichi Amaha}
\affiliation{Center for Emergent Matter Science, RIKEN, 2-1 Hirosawa, Wako, Saitama 351-0198, Japan}

\author{Jun Yoneda}
\affiliation{Center for Emergent Matter Science, RIKEN, 2-1 Hirosawa, Wako, Saitama 351-0198, Japan}
\affiliation{Department of Applied Physics, University of Tokyo, Bunkyo, Tokyo 113-8656, Japan}

\author{Kenta Takeda}
\affiliation{Center for Emergent Matter Science, RIKEN, 2-1 Hirosawa, Wako, Saitama 351-0198, Japan}

\author{Giles Allison}
\affiliation{Center for Emergent Matter Science, RIKEN, 2-1 Hirosawa, Wako, Saitama 351-0198, Japan}

\author{Sen Li}
\affiliation{Center for Emergent Matter Science, RIKEN, 2-1 Hirosawa, Wako, Saitama 351-0198, Japan}

\author{Akito Noiri}
\affiliation{Center for Emergent Matter Science, RIKEN, 2-1 Hirosawa, Wako, Saitama 351-0198, Japan}
\affiliation{Department of Applied Physics, University of Tokyo, Bunkyo, Tokyo 113-8656, Japan}

\author{Takumi Ito}
\affiliation{Center for Emergent Matter Science, RIKEN, 2-1 Hirosawa, Wako, Saitama 351-0198, Japan}
\affiliation{Department of Applied Physics, University of Tokyo, Bunkyo, Tokyo 113-8656, Japan}

\author{Daniel Loss}
\affiliation{Center for Emergent Matter Science, RIKEN, 2-1 Hirosawa, Wako, Saitama 351-0198, Japan}
\affiliation{Department of Physics, University of Basel, Klingelbergstrasse 82, 4056 Basel, Switzerland}

\author{Arne Ludwig}
\affiliation{Angewandte Festk\"orperphysik, Ruhr-Universit\"at Bochum, D-44780 Bochum, Germany}

\author{Andreas D. Wieck}
\affiliation{Angewandte Festk\"orperphysik, Ruhr-Universit\"at Bochum, D-44780 Bochum, Germany}

\author{Seigo Tarucha}%
\email[]{tarucha@ap.t.u-tokyo.ac.jp}
\affiliation{Center for Emergent Matter Science, RIKEN, 2-1 Hirosawa, Wako, Saitama 351-0198, Japan}
\affiliation{Department of Applied Physics, University of Tokyo, Bunkyo, Tokyo 113-8656, Japan}
\affiliation{Quantum-Phase Electronics Center, University of Tokyo, Bunkyo, Tokyo 113-8656, Japan}
\affiliation{Institute for Nano Quantum Information Electronics, University of Tokyo, 4-6-1 Komaba, Meguro, Tokyo 153-8505, Japan}

\date{\today}

\begin{abstract}
We analyze time evolution of charge and spin states in a quantum dot coupled to an electric reservoir.
Utilizing high-speed single-electron detection, we focus on dynamics induced by the first-order tunneling. 
We find that there is a difference between the spin and the charge relaxation: the former appears slower than the latter.
The difference depends on the Fermi occupation factor and the spin relaxation becomes slower when the energy level of the quantum dot is lowered.
We explain this behavior by a theory which includes the first-order tunneling processes.
We conduct detailed comparison of the experiment and the theory with changing the energy of the quantum dot levels, and the theory can reproduce the experimental results.
\end{abstract}

\maketitle


Semiconductor quantum dots (QDs) offer artificial quantum systems which can be controlled by voltages applied on gate electrodes~\cite{1996TaruchaPRL, 1997KouwenhovenSci, 2000CiorgaPRB, 2001RPP, 2002vanderWielRMP}.
By coupling the QDs to electronic reservoirs, we can explore the physics of quantum systems through electron transport measurements.
The electron tunneling through the QDs reflects internal levels of the QDs and the transport spectroscopy has been a key technique to probe the inner levels.
In addition, higher-order tunneling processes result in interesting physics of cotunneling~\cite{2001FranceschiPRL, 2005SchleserPRL} and the Kondo effect~\cite{1998GordonNat, 2000vanderWielSci}.
In recent years, sensitive high-speed transport measurement of QD systems became possible by utilizing quantum point contacts or QD charge sensors, and RF-reflectometry~\cite{1998SchelkopfSci, 2007ReillyAPL, 2010BarthelPRB}.
The technique is established and further developed in spin-based quantum bit experiments~\cite{1998LossPRA, 2008ReillySci, 2016OtsukaSciRep, 2018YonedaNatureNano} and realized fast qubit readout~\cite{2017NakajimaPRL} utilizing spin to charge conversion by Pauli spin blockade~\cite{2002OnoSci}.
The method is also useful to explore the dynamics of open quantum systems formed by QD-lead hybrid systems~\cite{2002KobayashiPRL, 2004ElzermanNature, 2007OtsukaJPSJ, 2008AmashaPRB, 2011SimmonsPRL, 2016HofmannPRL}.
We have previously demonstrated the measurement of charge and spin dynamics induced by the first- and higher-order tunneling processes, and revealed the time evolution by high-speed charge and spin measurements~\cite{2017OtsukaSciRep}.

In this work, we focus on the difference in the charge and spin dynamics in the first-order tunneling processes.
We observe that the spin equilibration is slower than the charge equilibration.
A theory treating the first-order tunneling explains the difference.
We conduct detailed comparison of the experiment and the theory with changing the energy of the QD levels against the lead's Fermi level.
The theory can reproduce the experimental results.


\begin{figure}
\begin{center}
  \includegraphics{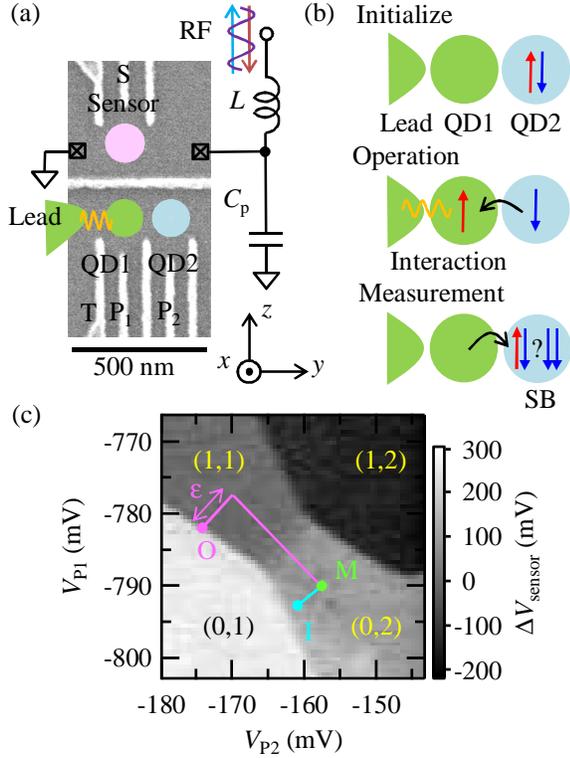}
  \caption{(a) SEM image of the device and the schematic of the measurement circuit.
  The target QD1 is connected to the ancillary QD2 for the spin initialization and readout.
  The charge state is monitored by the QD Sensor connected to the RF resonator circuit.
  (b) Schematics of the measurement procedure.
  The spin state is initialized in QD2 and the electron is transferred to QD1.
  After that, the charge and spin states evolve by the interaction with the lead.
  The charge state is monitored by the QD sensor.
  The final spin state after some interaction duration is measured utilizing the spin blockade. 
  (c) Charge stability diagram $\Delta V_{\rm sensor}$ as a function of $V_{\rm P2}$ and $V_{\rm P1}$.
  I, O and M correspond to gate voltage conditions for initialization, operation, and readout, respectively.
  The number of electrons in each QD is shown as $(n_{\rm 1}, n_{\rm 2})$.
  }
  \label{Device}
\end{center}
\end{figure}

Figure~\ref{Device}(a) shows a scanning electron micrography (SEM) image of the device and the schematic of the measurement circuit.
The device was fabricated from a GaAs/AlGaAs heterostructure wafer with an electron sheet carrier density of 2.0~$\times$~10$^{15}$~m$^{-2}$ and a mobility of 110~m$^2$/Vs at 4.2~K.
The two-dimensional electron gas is formed 90~nm under the wafer surface and patterned into QDs by applying negative voltages on Ti/Au Schottky surface gates which appear white in Fig.~\ref{Device}(a).
The target QD1 is connected to the lead through a tunneling barrier, which is controlled by gate T.
QD1 is also connected to the ancillary QD2 for spin initialization and readout~\cite{2005PettaSci, 2007HansonRMP}.
The charge state of QD1 and QD2 is monitored by a QD sensor formed at the upper side of the device.
The sensor is connected to an radio frequency (RF) tank circuit for the RF reflectometry and information of the charge state of the double QD (DQD) is extracted from the reflected RF signal~\cite{2010BarthelPRB, 2007ReillyAPL}.
All measurements were conducted in a dilution fridge cryostat.

Figure~\ref{Device}(b) shows a schematic of the measurement procedure: initialization, operation and measurement.
Figure~\ref{Device}(c) is the charge stability diagram showing the charge sensing signal $\Delta V_{\rm sensor}$ as a function of $V_{\rm P2}$ and $V_{\rm P1}$.
The number of electrons in each QD is shown as $(n_{\rm 1}, n_{\rm 2})$.
I, O, and M indicate the gate voltages corresponding to the initialization, operation, and measurement. 
The spin state is initialized in QD2 by utilizing the singlet formation at the configuration I.
After that, the initialized electron is transferred to QD1 by moving to the configuration O.
$\epsilon $ is the voltage from the center of the Coulomb blocked (1,1) charge configuration region.
As O is close to the charge transition line, the electron in QD1 interacts with the lead through the first-order tunneling processes.
The change of the charge state is monitored by the sensor during this phase.
The change of the spin state is deduced by the subsequent spin blockade measurement~\cite{2002OnoSci} at the configuration M.


\begin{figure}
\begin{center}
  \includegraphics{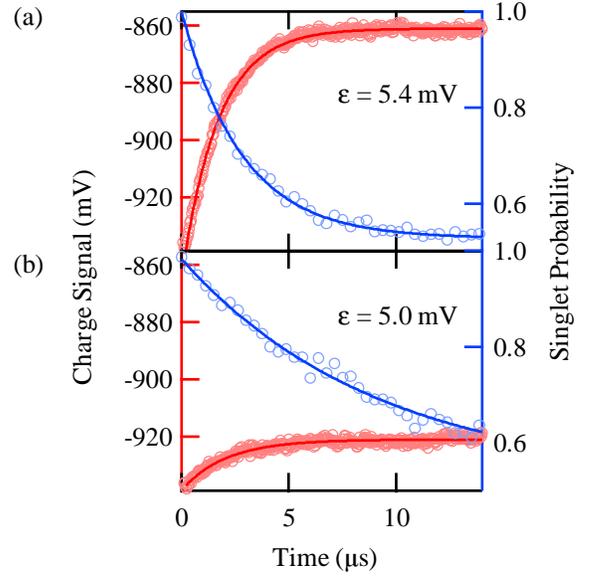}
  \caption{(a) Observed charge and spin signals as a function of the interaction time.
  Red circles show the charge signal (the average of the sensor signal $\langle V_{\rm sensor} \rangle $).
  Blue circles show the spin signal (the probability to find a singlet in M).
  The solid lines are exponential fits resulting in a relaxation time of 1.7~$\mu$s for the charge, and 3.0~$\mu$s for the spin.
  The QD level is close to the Fermi level of the reservoir ($\epsilon = 5.4$ mV).
  (b) Observed charge and spin signals as a function of the interaction time when we lowered the QD level ($\epsilon = 5.0$ mV).
  The spin relaxation time becomes slower, while the charge relaxation time is not affected.
  }
  \label{Relaxation}
\end{center}
\end{figure}

Figure~\ref{Relaxation}(a) shows the observed charge and spin signals as a function of the interaction time with the lead at O.
Red circles show the charge signal.
The signal is an average of 16384 traces by the repeated real-time charge detections during O $\langle V_{\rm sensor} \rangle $.
Blue circles show the spin signal (the singlet probability) measured at M after spending a fixed interaction time at O.
A single data point is extracted from 512 measurements.
Both the charge and spin signals change in time and show relaxation.
From the previous detailed measurement of the charge state~\cite{2017OtsukaSciRep}, we know that the mechanism of this relaxation is the first-order tunneling processes: the electron shuttles between the dot and the lead, and the charge and spin states change.
The relaxation time is determined by the tunneling rate, and the rate can be controlled by the voltage on gate T.
Note that the intrinsic spin relaxation time in a QD without electron tunneling (several hundreds of $\mu$s, ms)~\cite{2007HansonRMP} is much longer than this time scale.

The solid lines in Fig.~\ref{Relaxation}(a) are the result of the fitting with single exponential relaxation curves.
The charge relaxation time is 1.7~$\mu$s and the spin relaxation time is 3.0~$\mu$s. 
They are of the same order but still there is a difference: the charge relaxation is faster than the spin relaxation.
This goes against naive intuition that those should be the same since both the charge and the spin are carried by a single electron. 

This difference is enhanced when we lower the energy level of the QD against the Fermi level of the lead.
Figure~\ref{Relaxation}(b) is the result when we lowered the QD level by setting $\epsilon = 5.0$ mV.
The charge relaxation time is 2.1~$\mu$s and the spin relaxation time is 9.1~$\mu$s now.
The charge relaxation time is not significantly affected.
(The small change is induced by the change of the barrier height induced by the change of the operation point.)
On the other hand, the spin relaxation time becomes three times longer.


\begin{figure}
\begin{center}
  \includegraphics{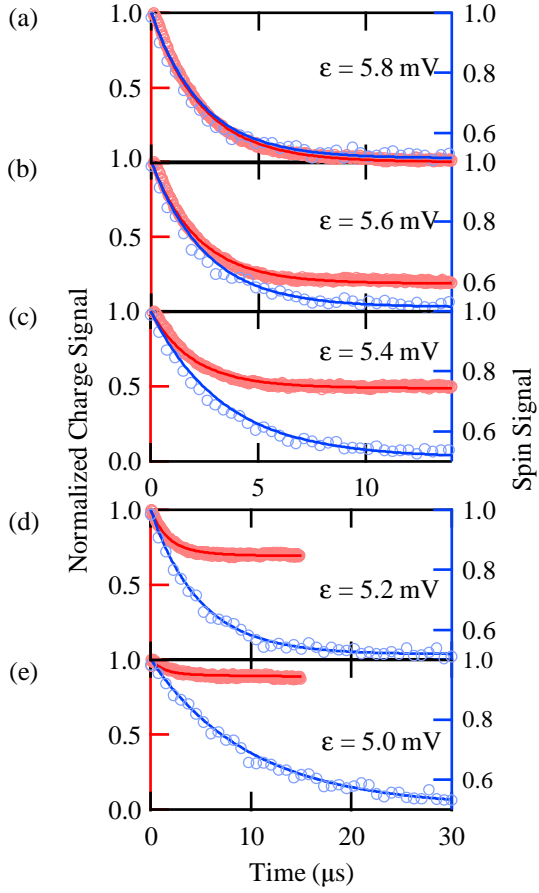}
  \caption{The observed charge (red) and spin (blue) signals for different energy alignment of the QD with respect to the Fermi level of the lead.
  The charge and spin signals are normalized.
  (a), (b), (c), (d) and (e) correspond to the results with $\epsilon = 5.8, 5.6, 5.4, 5.2, 5.0$ mV, respectively.
  }
  \label{Dependence}
\end{center}
\end{figure}

Figure~\ref{Dependence} shows a series of the observed charge and spin signals for different alignments of the QD levels with respect to the lead.
In these figures, the charge signal is normalized to span the range form zero to one.
(The normalization is done using the initial and saturated charge signal values at $f \approx 0$ and their dependence on $\epsilon $.)
(a), (b), (c), (d) and (e) correspond to the results with $\epsilon = 5.8, 5.6, 5.4, 5.2, 5.0$ mV, respectively.
When the QD level is above the Fermi level of the reservoir, so that the Fermi occupation factor $f \approx 0$, the time scales of the charge and the spin are almost same (Fig.~\ref{Dependence}(a)).
When we lower the QD level against the Fermi level of the lead, the spin relaxation becomes slower.
On the other hand, the charge relaxation timescale is not affected and the decay amplitude becomes smaller (Fig~\ref{Dependence}(b), (c), (d) and (e)).


\begin{figure}
\begin{center}
  \includegraphics{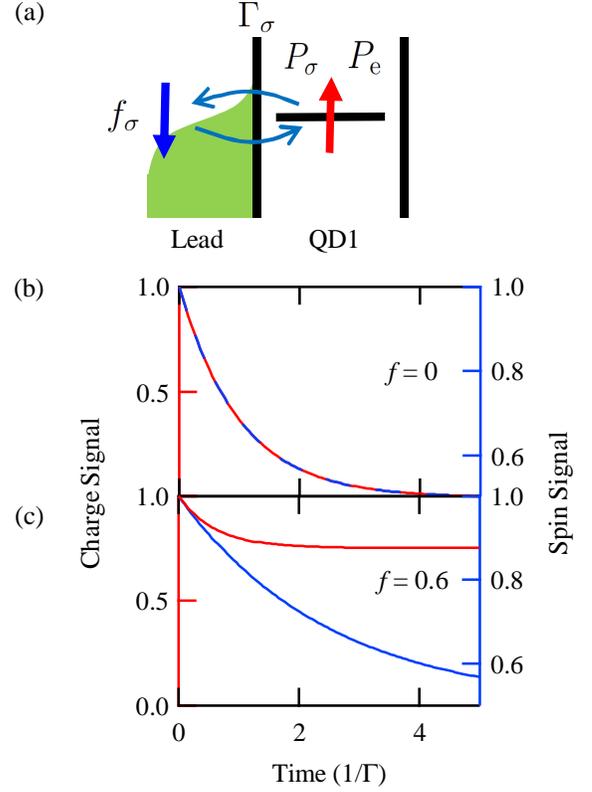}
  \caption{(a) Schematic of the theoretical model.
  The first-order tunneling event between the QD and the lead changes the probability of each state $P_\sigma, P_{\rm e}$.
  (b), (c) Calculated charge and spin dynamics using the model in (a) with experimental parameters (magnetic field $B=0.5$~T, electron temperature $T_{\rm e}=0.25$~K).
  (b) and (c) are results of $f=0, 0.6$ respectively.
  }
  \label{Theory}
\end{center}
\end{figure}

To reproduce the observed difference in the charge and spin relaxation and their QD energy level dependence, we set up a rate-equation model including the first-order tunneling processes~\cite{2017OtsukaSciRep}.
Figure~\ref{Theory}(a) shows a schematic of the theoretical model.
The first-order tunneling event between the QD and the lead changes the probabilities according to
\begin{equation}
\partial_t P_\sigma=-\Gamma_\sigma (1-f_\sigma)P_\sigma+\Gamma_\sigma f_\sigma P_{\rm e}.
\label{eq:Ps}
\end{equation}
Here, $P_\sigma$  is the probability that the dot is occupied by a single electron with spin $\sigma \in \{ \uparrow, \downarrow\}$, and $P_{\rm e}$ is the probability that the dot is empty. 
Further, $f_\sigma = f(\mu -\sigma g \mu _{\rm B}B/2)$ and $\Gamma_\sigma $ are the Fermi occupation factor and the tunneling rate for an electron with spin $\sigma $~\cite{2010PeterPRB}.
By solving this equation with $P_\uparrow+P_\downarrow+P_{\rm e}=1$, we can calculate the charge and spin dynamics of the system, given the experimental parameters (magnetic field $B=0.5$~T, electron temperature $T_{\rm e}=0.25$~K) and the initial condition $P_\uparrow (t=0)= 1$.

Figure~\ref{Theory}(b) and (c) shows calculated charge and spin dynamics at $f=0$ and 0.6, respectively.
Here, we assumed $\Gamma_\uparrow = \Gamma_ \downarrow = \Gamma $, for simplicity.
The charge relaxation time is not affected by the change of $f$ and only the decay amplitude decreases.
On the other hand, the spin relaxation time becomes longer with the increase of $f$.
These results are qualitatively the same to the observed experimental results in Fig.~\ref{Relaxation}.
The qualitative explanation of the difference is that the charge state of the QD is the result of both tunneling out and in process and then the time dependent terms including $f$ are canceled out.
On the other hand, the spin information is lost when the tunneling out process happens.
For a large value of $f$, the electron has to first tunnel out, which happens with a small rate $\propto (1-f)\Gamma$.
The smallness of the rate is the reason why spin relaxation looks slower than the charge relaxation.


The solid lines in Fig.~\ref{Dependence}(a), (b), (c), (d), and (e) show results of the fitting by the theoretical curves.
The charge and spin signal fitting share the same fitting parameters $f$ and $\Gamma $. 
The fitting parameters become $(f, \Gamma ({\rm MHz}))=(0.002, 0.42), (0.10, 0.43), (0.32, 0.41), (0.52, 0.41), (0.78, 0.46)$ for (a), (b), (c), (d), and (e), respectively.
With the decrease of $\epsilon $, the Fermi factor $f$ increases monotonically.
The theoretical fitting is consistent with the experimental data with reasonable fitting parameters.
This implies that our model captures the basic physics of the system induced by the first order tunneling.

In conclusion, we analyzed the difference in the charge and spin relaxation in a QD-lead hybrid system induced by first-order tunneling processes.
The difference depends on the Fermi occupation factor and the spin relaxation becomes slower when the energy level of the QD is lowered below the Fermi level of the lead.
A theory describing the first-order tunneling process reproduces the observed experimental results.
These results will be important for spin initializations and manipulations utilizing the coupling to the lead.


We thank K. Ono, RIKEN CEMS Emergent Matter Science Research Support Team for fruitful discussions and technical supports.
Part of this work is supported by 
ImPACT Program of Council for Science, Technology and Innovation (Cabinet Office, Government of Japan), 
Grants-in-Aid for Scientific Research (No. 26220710, 16H00817, 16K05411, 17H05187), 
CREST (JPMJCR15N2, JPMJCR1675), PRESTO (JPMJPR16N3), JST, 
RIKEN Incentive Research Project, 
The Telecommunications Advancement Foundation Research Grant, 
Futaba Electronics Memorial Foundation Research Grant, 
Kato Foundation for Promotion of Science Research Grant,
Hitachi Global Foundation Kurata Grant,
The Okawa Foundation for Information and Telecommunications Research Grant,
The Nakajima Foundation Research Grant,
Japan Prize Foundation Research Grant,
Iketani Science and Technology Foundation Research Grant,
Yamaguchi Foundation Research Grant,
DFG-TRR160, and the BMBF - Q.com-H  16KIS0109.

\end{document}